\documentclass{llncs}

\usepackage[linesnumbered,lined,boxed,vlined]{algorithm2e}
\usepackage{amsfonts}
\usepackage{amsmath}
\usepackage{arydshln}  
\usepackage[table]{xcolor}  
\usepackage{graphics}  
\usepackage{multirow}  
\def\b{\ensuremath\boldsymbol}
\usepackage{graphicx}

\usepackage{tikz}
\usepackage{lipsum}
\newcommand\copyrighttextt{%
  \footnotesize Published at International Conference on Image Analysis and Recognition, Springer. This version includes the supplementary material for derivation of some equations.}
\newcommand\copyrightnotice{%
\begin{tikzpicture}[remember picture,overlay]
\node[anchor=south,yshift=10pt] at (current page.south) {\fbox{\parbox{\dimexpr\textwidth-\fboxsep-\fboxrule\relax}{\copyrighttextt}}};
\end{tikzpicture}%
}

\begin{document}

\title{Principal Component Analysis \\Using Structural Similarity Index for Images}
\author{Benyamin Ghojogh,
Fakhri Karray,
Mark Crowley
}

\institute{Department of Electrical and Computer Engineering,\\ University of Waterloo, Waterloo, ON, Canada  \\
\email{\{bghojogh, karray, mcrowley\}@uwaterloo.ca}
}

\maketitle              

\begin{abstract}
Despite the advances of deep learning in specific tasks using images, the principled assessment of image fidelity and similarity is still a critical ability to develop.
As it has been shown that Mean Squared Error (MSE) is insufficient for this task, other measures have been developed with one of the most effective being Structural Similarity Index (SSIM). Such measures can be used for subspace learning but existing methods in machine learning, such as Principal Component Analysis (PCA), are based on Euclidean distance or MSE and thus cannot properly capture the structural features of images. In this paper, we define an image structure subspace which discriminates different types of image distortions. We propose Image Structural Component Analysis (ISCA) and also kernel ISCA by using SSIM, rather than Euclidean distance, in the formulation of PCA. This paper provides a bridge between image quality assessment and manifold learning opening a broad new area for future research. 
\end{abstract}

\copyrightnotice

\section{Introduction}

It has been shown that Mean Squared Error (MSE) is not a promising measure for image quality, fidelity, or similarity \cite{wang2009mean}. The distortions of an image or similarities of two images can be divided into two main categories, i.e., structural and non-structural distortions \cite{wang2004image}. The structural distortions, such as JPEG blocking distortion, Gaussian noise, and blurring, are the ones which are easily noticeable by Human Visual System (HVS), whereas the non-structural distortions, such as luminance enhancement and contrast change, do not have large impact on the visual quality of image. 

Structural similarity index (SSIM) \cite{wang2004image,wang2006modern} has been shown to be an effective measure for image quality assessment. It encounters luminance and contrast change as non-structural distortions and other distortions as structural ones. Due to its performance, it has recently been noticed and used in optimization problems \cite{brunet2018optimizing} for tasks such as image denoising, image restoration, contrast enhancement, image quantization, compression, etc, noticing that the distance based on SSIM is quasi-convex under certain conditions \cite{brunet2012mathematical}. 

So far, the fields of manifold learning and machine learning have largely used MSE and Euclidean distance in order to develop algorithms for subspace learning. Principal Component Analysis (PCA) is an example based on Euclidean distance or $\ell_2$ norm. However, MSE is not as promising as SSIM for image structure measurement \cite{wang2009mean,wang2006modern} making these algorithms not effective enough in terms of capturing the structural features of image. In this paper, we introduce the new concept of \textit{image structure subspace} which is a subspace capturing the intrinsic features of an image in terms of structural similarity and distortions, and can discriminate the various types of image distortions. This subspace can also be useful for parameter estimation for (or selection between) different denoising methods, but that topic will be dealt with in future work. 

The outline and contributions of the paper are as follows:
We begin by defining the background methods of SSIM and PCA. 
We then introduce ISCA using orthonomal bases and kernals by analogy to PCA, where ISCA can be seen as PCA which uses SSIM instead of the $\ell_2$ norm. We then describe an extensive set of experiments demonstrating the performance of ISCA on projection, reconstruction and out-of-sample analysis tasks compared to various kernel PCA methods.
The derivations of expressions in this paper are detailed more in the supplementary-material paper which will be released in \url{https://arXiv.org}.


\section{Structural Similarity Index}

The SSIM between two reshaped image blocks $\breve{\b{x}}_1 = [x_1^{(1)}, \dots, x_1^{(q)}]^\top \in \mathbb{R}^q$ and $\breve{\b{x}}_2 = [x_2^{(1)}, \dots, x_2^{(q)}]^\top \in \mathbb{R}^q$, in color intensity range $[0,l]$, is \cite{wang2004image,wang2006modern}:
\begin{align}
\mathbb{R} \ni \text{SSIM}(\breve{\b{x}}_1, \breve{\b{x}}_2) := \bigg(\frac{2 \mu_{x_1} \mu_{x_2} + c_1}{\mu_{x_1}^2 + \mu_{x_2}^2 + c_1}\bigg) \bigg(\frac{2 \sigma_{x_1} \sigma_{x_2} + c_2}{\sigma_{x_1}^2 + \sigma_{x_2}^2 + c_2}\bigg) \bigg(\frac{\sigma_{x_1, x_2} + c_3}{\sigma_{x_1}\sigma_{x_2} + c_3}\bigg),
\end{align}
where $\mu_{x_1} = (1/q) \sum_{i=1}^q x_1^{(i)}$, $\sigma_{x_1} = \Big[\big(1/(q-1)\big) \sum_{i=1}^q (x_1^{(i)} - \mu_{x_1})^2\Big]^{0.5}$, $\sigma_{x_1,x_2} = \big(1/(q-1)\big) \sum_{i=1}^q (x_1^{(i)} - \mu_{x_1}) (x_2^{(i)} - \mu_{x_2})$, $c_1=(0.01 \times l)^2$, $c_2=2\,c_3=(0.03 \times l)^2$, and $\mu_{x_2}$ and $\sigma_{x_2}$ are defined similarly for $\breve{\b{x}}_2$. 
In this work, $l=1$.
The $c_1$, $c_2$, and $c_3$ are for avoidance of singularity \cite{wang2006modern} and $q$ is the dimensionality of the reshaped image patch.
Note that since $c_2=2\,c_3$, we can simplify SSIM to $\text{SSIM}(\breve{\b{x}}_1, \breve{\b{x}}_2) = s_1(\breve{\b{x}}_1, \breve{\b{x}}_2) \times s_2(\breve{\b{x}}_1, \breve{\b{x}}_2)$, where $s_1(\breve{\b{x}}_1, \breve{\b{x}}_2) := (2 \mu_{x_1} \mu_{x_2} + c_1)/(\mu_{x_1}^2 + \mu_{x_2}^2 + c_1)$ and $s_2(\breve{\b{x}}_1, \breve{\b{x}}_2) := (2 \sigma_{x_1, x_2} + c_2)/(\sigma_{x_1}^2 + \sigma_{x_2}^2 + c_2)$.
If the vectors $\breve{\b{x}}_1$ and $\breve{\b{x}}_2$ have zero mean, i.e., $\mu_{x_1} = \mu_{x_2} = 0$, the SSIM becomes $\mathbb{R} \ni \text{SSIM}(\breve{\b{x}}_1, \breve{\b{x}}_2) = (2\breve{\b{x}}_1^\top \breve{\b{x}}_2 + c) / (||\breve{\b{x}}_1||_2^2 + ||\breve{\b{x}}_2||_2^2 + c)$, where $c = (q-1) \,c_2$ \cite{otero2014unconstrained}.
We denote the reshaped vectors of the two images by $\b{x}_1 \in \mathbb{R}^{d}$ and $\b{x}_2 \in \mathbb{R}^{d}$, and a reshaped block in the two images by $\breve{\b{x}}_1 \in \mathbb{R}^{q}$ and $\breve{\b{x}}_2 \in \mathbb{R}^{q}$.
The (squared) distance based on SSIM, which we denote by $||.||_S$, is \cite{otero2014unconstrained,brunet2012mathematical,brunet2011class}:
\begin{align}\label{equation_SSIM_distance}
\mathbb{R} \ni ||\breve{\b{x}}_1 - \breve{\b{x}}_2||_S := 1 -  \text{SSIM}(\breve{\b{x}}_1, \breve{\b{x}}_2) = \frac{||\breve{\b{x}}_1 - \breve{\b{x}}_2||_2^2}{||\breve{\b{x}}_1||_2^2 + ||\breve{\b{x}}_2||_2^2 + c},
\end{align}
where $\mu_{x_1} = \mu_{x_2} = 0$.
In ISCA and PCA which inspires ISCA, the data should be centered; therefore, the fact that $\breve{\b{x}}_1$ and $\breve{\b{x}}_2$ should be centered is useful.

\section{Principal Component Analysis}

Since ISCA is inspired by PCA \cite{jolliffe2011principal} we briefly review it here.
Assume that the orthonormal columns of matrix $\b{U} \in \mathbb{R}^{d \times p}$ are the vectors which span the PCA subspace. Then, the projected data $\widetilde{\b{X}} \in \mathbb{R}^{p \times n}$ onto PCA subspace and the reconstructed data $\hat{\b{X}} \in \mathbb{R}^{d \times n}$ are $\widetilde{\b{X}} = \b{U}^\top \b{X}$ and $\hat{\b{X}} = \b{U} \widetilde{\b{X}} = \b{U} \b{U}^\top \b{X}$, respectively.
The squared length of the projected data is $||\hat{\b{X}}||_F^2 = ||\b{U}\b{U}^\top \b{X}||_F^2 = \textbf{tr}(\b{U}^\top \b{X} \b{X}^\top \b{U})$
where $\textbf{tr}(.)$ and $||.||_F$ denote the trace and Frobenius norm of matrix, respectively.
Presuming that the data $\b{X}$ are already centered, the $\b{S} = \b{X} \b{X}^\top$ is the covariance matrix; therefore: $||\hat{\b{X}}||_F^2 = \textbf{tr}(\b{U}^\top \b{S}\, \b{U})$. Maximizing the squared length of projection where the projection matrix is orthogonal is:
\begin{equation}
\begin{aligned}
& \underset{\b{U}}{\text{maximize}}
& & \textbf{tr}(\b{U}^\top \b{S}\, \b{U}), \\
& \text{subject to}
& & \b{U}^\top \b{U} = \b{I},
\end{aligned}
\end{equation}
The Lagrangian \cite{boyd2004convex} is: $\mathcal{L} = \textbf{tr}(\b{U}^\top \b{S}\, \b{U}) - \textbf{tr}\big(\b{\Lambda}^\top (\b{U}^\top \b{U} - \b{I})\big)$, where $\b{\Lambda} \in \mathbb{R}^{p \times p}$ is a diagonal matrix including Lagrange multipliers. Equating derivative of $\mathcal{L}$ to zero gives us: $\mathbb{R}^{d \times p} \ni \partial \mathcal{L} / \partial \b{U} = 2\,\b{S} \b{U} - 2\,\b{U} \b{\Lambda} \overset{\text{set}}{=} \b{0} \implies \b{S} \b{U} = \b{U} \b{\Lambda}$.
Therefore, columns of $\b{U}$ are the eigenvectors of the covariance matrix $\b{S}$.

PCA can be looked at with another point of view.
The reconstruction error is $\b{R} := \b{X} - \hat{\b{X}} = \b{X} - \b{U} \b{U}^\top \b{X}$ where $\mathbb{R}^{d \times n} \ni \b{R} = [\b{r}_1, \dots, \b{r}_n]$ is the matrix of residuals.
We want to minimize the reconstruction error:
\begin{equation}\label{equation_optimization_reconstruction}
\begin{aligned}
& \underset{\b{U}}{\text{minimize}}
& & ||\b{X} - \b{U}\b{U}^\top\b{X}||_F^2, \\
& \text{subject to}
& & \b{U}^\top \b{U} = \b{I}.
\end{aligned}
\end{equation}
The objective function is $||\b{X} - \b{U}\b{U}^\top\b{X}||_F^2 = \textbf{tr}(\b{X}^\top\b{X}-\b{X}\b{X}^\top\b{U}\b{U}^\top)$.
The Lagrangian \cite{boyd2004convex} is: $\mathcal{L} = \textbf{tr}(\b{X}^\top\b{X})-\textbf{tr}(\b{X}\b{X}^\top\b{U}\b{U}^\top) - \textbf{tr}\big(\b{\Lambda}^\top (\b{U}^\top \b{U} - \b{I})\big)$, where $\b{\Lambda} \in \mathbb{R}^{p \times p}$ is a diagonal matrix including Lagrange multipliers. Equating the derivative of $\mathcal{L}$ to zero gives: $\partial \mathcal{L} / \partial \b{U} = 2\,\b{X}\b{X}^\top \b{U} - 2\,\b{U} \b{\Lambda} \overset{\text{set}}{=} \b{0} \implies \b{X}\b{X}^\top \b{U} = \b{U} \b{\Lambda} \implies \b{S} \b{U} = \b{U} \b{\Lambda}$, which is again the eigenvalue problem for the covariance matrix $\b{S}$. Therefore, PCA subspace is the best linear projection in terms of reconstruction error.

As shown above, PCA \cite{jolliffe2011principal} is based on $\ell_2$ (or Frobenius) norm which is not a promising measure for image quality assessment \cite{wang2009mean}. In order to have both the minimization of reconstruction error as in PCA and using a proper measure for image fidelity, we propose ISCA.

\section{Image Structural Component Analysis (ISCA)}

\subsection{Orthonormal Bases for One Image}

Our goal is to find a subspace spanned by $p$ directions for some desired $p$.
Consider an image block $\breve{\b{x}} \in \mathbb{R}^q$ which is centered (its mean is removed). We want to project it onto a $p$-dimensional subspace and then reconstruct it back, where $p \leq q$. Assume $\mathbb{R}^{q \times p} \ni \b{U} := [\b{u}_1, \dots, \b{u}_p]$ is a matrix whose columns are the projection directions spanning the subspace. The projection and reconstruction of $\breve{\b{x}}$ are $\b{U}^\top \breve{\b{x}}$ and $\b{U} \b{U}^\top \breve{\b{x}}$, respectively.
We want to minimize the reconstruction error with orthonormal bases of the subspace; therefore:
\begin{equation}\label{equation_optimization_ISCA_oneBlock}
\begin{aligned}
& \underset{\b{U} \in \mathbb{R}^{q \times p}}{\text{minimize}}
& & ||\breve{\b{x}} - \b{U}\b{U}^\top\breve{\b{x}}||_S, \\
& \text{subject to}
& & \b{U}^\top \b{U} = \b{I}.
\end{aligned}
\end{equation}
According to Eq. (\ref{equation_SSIM_distance}) and noticing the orthonormality of projection directions, $\b{U}^\top\b{U} = \b{I}$, we have:
\begin{align}\label{equation_f_U}
\mathbb{R} \ni f(\b{U}) := ||\breve{\b{x}} - \b{U}\b{U}^\top\breve{\b{x}}||_S = \frac{\breve{\b{x}}^\top (\b{I} - \b{U}\b{U}^\top)\, \breve{\b{x}}}{\breve{\b{x}}^\top (\b{I} + \b{U}\b{U}^\top)\, \breve{\b{x}} + c}.
\end{align}
The gradient of the $f(\b{U})$ is:
\begin{align}\label{equation_gradient_of_f}
\mathbb{R}^{q \times p} \ni \b{G}(\b{U}) := \frac{\partial f(\b{U})}{\partial\, \b{U}} = \frac{-2\,(1 + f(\b{U}))}{||\breve{\b{x}}||_2^2 + ||\b{U}\b{U}^\top\breve{\b{x}}||_2^2 + c} \,\breve{\b{x}}\breve{\b{x}}^\top\b{U}.
\end{align}

We partition a $d$-dimensional image into $b = \lceil d / q \rceil$ non-overlapping blocks each of which is a reshaped vector $\breve{\b{x}} \in \mathbb{R}^q$. The parameter $q$ is an upper bound on the desired dimensionality of the subspace of block ($p \leq q$). This parameter should not be a very large number due to the spatial variety of image statistics, yet also not very small so as to be able to capture the image structure. Also note that $p$ is an upper bound on the rank of $\b{U}\b{U}^\top \in \mathbb{R}^{q \times q}$.

We have $b$ instances of $p$-dimensional subspaces, one for each of the blocks. For projecting an image into the subspace and reconstructing it back, one can project and reconstruct every block of an image separately using the $p$ bases of the block subspace. The overall bases of an image can be visualized in image-form by putting the bases of blocks next to each other (see the experiments in Section \ref{sec:experiments}).

Considering all the $b$ blocks in an image, the problem in Eq. (\ref{equation_optimization_ISCA_oneBlock}) becomes:
\begin{equation}\label{equation_optimization_ISCA_allBlocks}
\begin{aligned}
& \underset{\b{U}_i \in \mathbb{R}^{q \times p}}{\text{minimize}}
& & \sum_{i=1}^b ||\breve{\b{x}}_i - \b{U}_i\b{U}_i^\top\breve{\b{x}}_i||_S, \\
& \text{subject to}
& & \b{U}_i^\top \b{U}_i = \b{I}, ~~~ \forall i \in \{1, \dots, b\},
\end{aligned}
\end{equation}
where $\b{x}_i \in \mathbb{R}^q$ and $\b{U}_i \in \mathbb{R}^{q \times p}$ are the $i$-th block and the bases of its subspace, respectively.
We can embed the constraint as an indicator function in the objective function \cite{boyd2011distributed}: 
\begin{equation}\label{equation_ADMM_problem}
\begin{aligned}
& \underset{\b{U}_i, \b{V}_i \in \mathbb{R}^{q \times p}}{\text{minimize}}
& & \sum_{i=1}^b \big( f(\b{U}_i) + h(\b{V}_i) \big), \\
& \text{subject to}
& & \b{U} - \b{V} = \b{0}, 
\end{aligned}
\end{equation}
where $f(\b{U}_i) := ||\breve{\b{x}}_i - \b{U}_i\b{U}_i^\top\breve{\b{x}}_i||_S$ and $h(\b{V}_i) := \mathbb{I}(\b{V}_i^\top \b{V}_i = \b{I})$. The $\mathbb{I}(.)$ denotes the indicator function which is zero if its condition is satisfied and is infinite otherwise. The $\b{U}$ and $\b{V}$ are defined as union of partitions to form an image-form array, i.e., $\b{U} := \cup_{i=1}^b \b{U}_i$ and $\b{V} := \cup_{i=1}^b \b{V}_i$ \cite{otero2018alternate}.

The Eq. (\ref{equation_ADMM_problem}) can be solved using Alternating Direction Method of Multipliers (ADMM) \cite{boyd2011distributed,otero2018alternate}.
The augmented Lagrangian for Eq. (\ref{equation_ADMM_problem}) is: $\mathcal{L}_{\rho} = \sum_{i=1}^b \big( f(\b{U}_i) + h(\b{V}_i) \big) + \textbf{tr}\big(\b{\Lambda}^\top (\b{U} - \b{V})\big) + (\rho/2)\, ||\b{U} - \b{V}||_F^2 = \sum_{i=1}^b \big( f(\b{U}_i) + h(\b{V}_i) \big) + (\rho/2)\, ||\b{U} - \b{V} + \b{J}||_F^2 - (\rho/2)\, ||\b{\Lambda}||_F^2$, where $\b{\Lambda} := \cup_{i=1}^b \b{\Lambda}_i$ is the Lagrange multiplier, $\rho > 0$ is a parameter, and $\b{J} := (1/\rho) \b{\Lambda} = (1/\rho) \cup_{i=1}^b \b{\Lambda}_i = \cup_{i=1}^b \b{J}_i$. Note that the term $(\rho/2)\, ||\b{\Lambda}||_F^2$ is a constant with respect to $\b{U}$ and $\b{V}$ and can be dropped. 
The updates of $\b{U}$, $\b{V}$, and $\b{J}$ are done as \cite{boyd2011distributed,otero2018alternate}:
\begin{align}
\b{U}_i^{(k+1)} & := \arg \min_{\b{U}_i} \Big( f(\b{U}_i) + (\rho/2)\, ||\b{U}_i - \b{V}_i^{(k)} + \b{J}_i^{(k)}||_F^2 \Big), \label{equation_ADMM_U_update} \\
\b{V}_i^{(k+1)} & := \arg \min_{\b{V}_i} \Big( h(\b{V}_i) + (\rho/2)\, ||\b{U}_i^{(k+1)} - \b{V}_i + \b{J}_i^{(k)}||_F^2 \Big), \label{equation_ADMM_V_update} \\
\b{J}^{(k+1)} & := \b{J}^{(k)} + \b{U}^{(k+1)} - \b{V}^{(k+1)}. \label{equation_ADMM_J_update}
\end{align}

Considering $||\b{A}||_F^2 = \textbf{tr}(\b{A}^\top \b{A})$ for a matrix $\b{A}$, the gradient of the objective function in Eq. (\ref{equation_ADMM_U_update}) with respect to $\b{U}_i$ is $\b{G}(\b{U}_i) + \rho\, (\b{U}_i - \b{V}_i^{(k)} + \b{J}_i^{(k)})$ where $\b{G}(\b{U}_i)$ is defined in Eq. (\ref{equation_gradient_of_f}). We can use the gradient decent method \cite{boyd2004convex} for solving the Eq. (\ref{equation_ADMM_U_update}). Our experiments showed that even one iteration of gradient decent suffices for Eq. (\ref{equation_ADMM_U_update}) because the ADMM itself is iterative. Hence, we can replace this equation with one iteration of gradient decent. 

The proximal operator is defined as \cite{parikh2014proximal}:
\begin{align}\label{equation_prox_vector}
\textbf{prox}_{\lambda, h}(\b{v}) := \arg \min_{\b{u}}\,\big(h(\b{u}) + (\lambda/2) ||\b{u} - \b{v}||_2^2\big),
\end{align}
where $\lambda$ is the proximal parameter and $h$ is the function that the proximal algorithm wants to minimize.
According to Eq. (\ref{equation_prox_vector}), the Eq. (\ref{equation_ADMM_V_update}) is equivalent to $\textbf{prox}_{\rho, h}(\b{U}_i^{(k+1)} + \b{J}_i^{(k)})$. 
As $h(.)$ is indicator function, its proximal operator is projection \cite{parikh2014proximal}. Therefore, Eq. (\ref{equation_ADMM_V_update}) is equivalent to $\mathrm{\Pi}(\b{U}_i^{(k+1)} + \b{J}_i^{(k)})$ where $\mathrm{\Pi}(.)$ denotes projection onto a set. 
Here, the variable of proximal operator is a matrix and not a vector. 
According to \cite{parikh2014proximal}, if $F$ is a convex and orthogonally invariant function, and it works on the singular values of a matrix variable $\b{A} \in \mathbb{R}^{q \times p}$, i.e., $F = f \circ \sigma$ where the function $\sigma(\b{A})$ gives the vector of singular values of $\b{A}$, then the proximal operator is:
\begin{align}\label{equation_prox_matrix}
\textbf{prox}_{\lambda, F}(\b{A}) := \b{Q}\,\, \textbf{diag}\Big(\textbf{prox}_{\lambda, f}\big(\sigma(\b{A})\big)\Big)\,\, \b{\Omega}^\top.
\end{align}
The $\b{Q} \in \mathbb{R}^{q \times p}$ and $\b{\Omega} \in \mathbb{R}^{p \times p}$ are the matrices of left and right singular vectors of $\b{A}$, respectively.
In our constraint $\b{V}^\top \b{V} = \b{I}$, the function $F$ deals with the singular values of $\b{V}$. The reason is that we want: $\b{V} \overset{\text{SVD}}{=} \b{Q} \b{\Sigma} \b{\Omega}^\top \implies \b{V}^\top \b{V} = \b{\Omega} \b{\Sigma} \b{Q}^\top \b{Q} \b{\Sigma} \b{\Omega}^\top \overset{(a)}{=} \b{\Omega} \b{\Sigma}^2 \b{\Omega}^\top \overset{\text{set}}{=} \b{I} \implies \b{\Omega} \b{\Sigma}^2 \b{\Omega}^\top \b{\Omega} = \b{\Omega} \overset{(b)}{\implies} \b{\Omega} \b{\Sigma}^2 = \b{\Omega} \implies \b{\Sigma} = \b{I}$, where $(a)$ and $(b)$ are because $\b{Q}$ and $\b{\Omega}$ are orthogonal matrices. 
Therefore, we can use Eq. (\ref{equation_prox_matrix}) for Eq. (\ref{equation_ADMM_V_update}) where $\textbf{prox}_{\rho, h}(\b{U}_i^{(k+1)} + \b{J}_i^{(k)})$ sets the singular values of $(\b{U}_i^{(k+1)} + \b{J}_i^{(k)})$ to one. 
In summary, Eqs. (\ref{equation_ADMM_U_update}), (\ref{equation_ADMM_V_update}), and (\ref{equation_ADMM_J_update}) can be restated as:
\begin{equation}\label{equation_ADMM_update_2}
\begin{aligned}
\b{U}_i^{(k+1)} & := \b{U}_i^{(k)} - \eta\, \b{G}(\b{U}_i^{(k)}) - \eta\,\rho\, (\b{U}_i^{(k)} - \b{V}_i^{(k)} + \b{J}_i^{(k)}), \\
\b{V}_i^{(k+1)} & := \b{Q}_i\,\, \textbf{diag}\Big(\textbf{prox}_{\rho, h}\big(\sigma(\b{U}_i^{(k+1)} + \b{J}_i^{(k)})\big)\Big)\,\, \b{\Omega}_i^\top, \\
\b{J}^{(k+1)} & := \b{J}^{(k)} + \b{U}^{(k+1)} - \b{V}^{(k+1)},
\end{aligned}
\end{equation}
where columns of $\b{Q}_i \in \mathbb{R}^{q \times p}$ and $\b{\Omega}_i \in \mathbb{R}^{p \times p}$ are the left and right singular vectors of $(\b{U}_i^{(k+1)} + \b{J}_i^{(k)})$ and $\eta > 0$ is the learning rate.
Iteratively solving Eq. (\ref{equation_ADMM_update_2}) until convergence gives us the $\b{U}_i$ for for the image blocks indexed by $i$. The $p$ columns of $\b{U}_i$ are the bases for the \textit{ISCA subspace} of the $i$-th block. Unlike in PCA, the ISCA bases do not have an order of importance but as in PCA, they are orthogonal capturing different features of image structure.
The $i$-th projected block is $\b{U}_i^\top \breve{\b{x}}_i \in \mathbb{R}^p$ where its dimensions are \textit{image structural components}. Note that $\breve{\b{x}}_i$, whether it is a block in a training image or an out-of-sample image, is centered.
It is noteworthy that if we consider only one block in the images, the subscript $i$ is dropped from Eq. (\ref{equation_ADMM_update_2}).

\subsection{Orthonormal Bases for a Set of Images}\label{section_oneBlock_allImages}

So far, if we have a set of $n$ images, we can find the subspace bases $\b{U}_i$ for the $i$-th block in each of them using Eq. (\ref{equation_ADMM_update_2}).
Now, we want to find the subspace bases $\b{U}_i$ for the $i$-th block in \textit{all training images of the dataset}. In other words, we want to find the subspace for the best reconstruction of the $i$-th block in all training images.
For this goal, we can look at the optimization problem in Eq. (\ref{equation_optimization_ISCA_allBlocks}) or (\ref{equation_ADMM_problem}) as an undercomplete auto-encoder neural network \cite{goodfellow2016deep} with one hidden layer where the input layer, hidden layer, and output layer have $q$, $p$, and $q$ neurons, respectively. The $\b{U}_i^\top \breve{\b{x}}$ and $\b{U}_i \b{U}_i^\top \breve{\b{x}}$ fill the role of applying the first and second weight matrices to the input, respectively. The weights are $\b{U}_i \in \mathbb{R}^{q \times p}$. Therefore, we will have $b$ auto-encoders, each with one hidden layer.

For training the auto-encoder, we introduce the blocks in an image as the input to this network and update the weights $\b{U}_i, \forall i$ based on Eq. (\ref{equation_ADMM_update_2}). Note that we do this update of weights with only `one' iteration of ADMM. Then, we move to the blocks in the next image and update the weights $\b{U}_i, \forall i$ again by an iteration of Eq. (\ref{equation_ADMM_update_2}). We do this for all images one by one until an epoch is completed where an epoch is defined as introducing the block in all training images of dataset to the network.
After termination of an epoch, we start another epoch to tune the weights $\b{U}_i, \forall i$ again. The epochs are repeated until the convergence. 
The termination criterion can be average reconstruction error $(1/(nb)) \sum_{i=1}^b \sum_{j=1}^n ||\breve{\b{x}}_{j,i} - \b{U}_i\b{U}_i^\top\breve{\b{x}}_{j,i}||_S < \varepsilon$, where $\varepsilon$ is a small number and $\breve{\b{x}}_{j,i}$ is the $i$-th block in the $j$-th image.
After training the network, we have one $p$-dimensional subspace for every block in all training images where the columns of the weight matrix $\b{U}_i$ span the subspace.
Note that because of ADMM, the auto-encoders are trained simultaneously and in parallel. 
Again, the $p$ columns of $\b{U}_i$ are the bases for the \textit{ISCA subspace} of the $i$-th block. 

\section{Kernel Image Structural Component Analysis}

We can map the block $\breve{\b{x}} \in \mathbb{R}^q$ to higher-dimensional feature space hoping to have the data fall close to a simpler-to-analyze manifold in the feature space. Suppose $\b{\phi}: \breve{\b{x}} \rightarrow \mathcal{H}$ is a function which maps the data $\breve{\b{x}}$ to the feature space. In other words, $\breve{\b{x}} \mapsto \b{\phi}(\breve{\b{x}})$. Let $t$ denote the dimensionality of the feature space, i.e., $\b{\phi}(\breve{\b{x}}) \in \mathbb{R}^t$. We usually have $t \gg q$. The kernel of the $i$-th block in images $1$ and $2$, which are $\breve{\b{x}}_{1,i}$ and $\breve{\b{x}}_{2,i}$, is $\b{\phi}(\breve{\b{x}}_{1,i})^\top \b{\phi}(\breve{\b{x}}_{2,i}) \in \mathbb{R}$ \cite{hofmann2008kernel}. The kernel matrix for the $i$-th block among the $n$ images is $\mathbb{R}^{n \times n} \ni \b{K}_i := \b{\Phi}(\breve{\b{X}}_i)^\top \b{\Phi}(\breve{\b{X}}_i)$ where $\b{\Phi}(\breve{\b{X}}_i) := [\b{\phi}(\breve{\b{x}}_{1,i}), \dots, \b{\phi}(\breve{\b{x}}_{n,i})] \in \mathbb{R}^{t \times n}$.
After calculating the kernel matrix, we normalize it \cite{ah2010normalized} as $\b{K}_i(a,b) := \b{K}_i(a,b) / \sqrt{\b{K}_i(a,a) \b{K}_i(b,b)}$ where $\b{K}_i(a,b)$ denotes the $(a,b)$-th element of the kernel matrix. Afterwards, the kernel is double-centered as $\b{K}_i := \b{H} \b{K}_i \b{H}$ where $\mathbb{R}^{n \times n} \ni \b{H} := \b{I} - (1/n) \b{1}\b{1}^\top$. 
The reason for double-centering is that Eq. (\ref{equation_SSIM_distance}) requires $\b{\phi}(\breve{\b{x}}_i)$ and thus the $\b{\Phi}(\breve{\b{X}}_i)$ to be centered (see Eq. (\ref{equation_optimization_kernel_ISCA_allBlocks})). 
Therefore, in kernel ISCA, we center the kernel rather than centering $\breve{\b{x}}$.

According to representation theory \cite{alperin1993local}, the projection matrix can be expressed as a linear combination of the projected data points. Therefore, we have $\mathbb{R}^{t \times p} \ni \b{\Phi}(\b{U}_i) = \b{\Phi}(\breve{\b{X}}_i)\, \b{\Theta}_i$ where every column of $\b{\Theta}_i := [\b{\theta}_1, \dots, \b{\theta}_p] \in \mathbb{R}^{n \times p}$ is the vector of coefficients for expressing a projection direction as a linear combination of projected image blocks.

As we did for ISCA, first we consider learning the $b$ subspaces for `one' image, here.
Considering $\b{\Phi}(\b{U}_i) = \b{\Phi}(\breve{\b{X}}_i)\, \b{\Theta}_i$ for the $i$-th block in the image, the objective function of Eq. (\ref{equation_optimization_ISCA_allBlocks}) in feature space is $\sum_{i=1}^b ||\b{\phi}(\breve{\b{x}}_i) - \b{\Phi}(\breve{\b{X}}_i)\, \b{\Theta}_i\, \b{\Theta}_i^\top \b{k}_i||_S$ where $\mathbb{R}^n \ni \b{k}_i := \b{\Phi}(\breve{\b{X}}_i)^\top \b{\phi}(\breve{\b{x}}_i)$.
Note that $\b{\Phi}(\breve{\b{X}}_i)$ includes the mapping of the $i$-th block in all the $n$ images while $\b{\phi}(\breve{\b{x}}_i)$ is the mapping of the $i$-th block in the image we are considering.
The constraint of Eq. (\ref{equation_optimization_ISCA_allBlocks}) in the feature space is $\b{\Phi}(\b{U}_i)^\top \b{\Phi}(\b{U}_i) = \b{\Theta}_i^\top \b{K}_i \b{\Theta}_i = \b{I}$.
Therefore, the Eq. (\ref{equation_optimization_ISCA_allBlocks}) in the feature space is:
\begin{equation}\label{equation_optimization_kernel_ISCA_allBlocks}
\begin{aligned}
& \underset{\b{\Theta}_i \in \mathbb{R}^{n \times p}}{\text{minimize}}
& & \sum_{i=1}^b ||\b{\phi}(\breve{\b{x}}_i) - \b{\Phi}(\breve{\b{X}}_i)\, \b{\Theta}_i\, \b{\Theta}_i^\top \b{k}_i||_S, \\
& \text{subject to}
& & \b{\Theta}_i^\top \b{K}_i\, \b{\Theta}_i = \b{I}, ~~~ \forall i \in \{1, \dots, b\}.
\end{aligned}
\end{equation}

Noticing the constraint $\b{\Theta}_i^\top \b{K}_i\, \b{\Theta}_i = \b{I}$ and using Eq. (\ref{equation_SSIM_distance}), we have:
\begin{align}\label{equation_f_kernel}
\mathbb{R} \ni f(\b{\Theta}_i) := ||\b{\phi}(\breve{\b{x}}_i) - \b{\Phi}(\breve{\b{X}}_i)\, \b{\Theta}_i\, \b{\Theta}_i^\top \b{k}_i||_S = \frac{k_i - \b{k}_i^\top \b{\Theta}_i \b{\Theta}_i^\top \b{k}_i}{k_i + \b{k}_i^\top \b{\Theta}_i \b{\Theta}_i^\top \b{k}_i + c},
\end{align}
where $\mathbb{R} \ni k_i := \b{\phi}(\breve{\b{x}}_i)^\top \b{\phi}(\breve{\b{x}}_i)$.
The gradient of the $f(\b{\Theta}_i)$ is:
\begin{align}\label{equation_gradient_of_f_kernel}
\mathbb{R}^{n \times p} \ni \b{G}(\b{\Theta}_i) := \frac{\partial f(\b{\Theta}_i)}{\partial\, \b{\Theta}_i} = \frac{-2\,(1 + f(\b{\Theta}_i))}{k_i + \b{k}_i^\top \b{\Theta}_i \b{\Theta}_i^\top \b{k}_i + c} \,\b{k}_i \b{k}_i^\top\b{\Theta}_i.
\end{align}

We can simplify the constraint $\b{\Theta}_i^\top \b{K}_i\, \b{\Theta}_i = \b{I}$. As the kernel $\b{K}_i$ is positive semi-definite, we can decompose it as: 
\begin{align*}
\mathbb{R}^{n \times n} \ni \b{K}_i \overset{\text{SVD}}{=} \b{\Psi} \b{\Upsilon} \b{\Psi}^\top = \b{\Psi} \b{\Upsilon}^{(1/2)}\b{\Upsilon}^{(1/2)} \b{\Psi}^\top = \b{\Delta}^\top \b{\Delta},
\end{align*}
where $\mathbb{R}^{n \times n} \ni \b{\Delta} := \b{\Upsilon}^{(1/2)} \b{\Psi}^\top$.
Therefore, the constraint can be written as: $\b{\Theta}_i^\top \b{K}_i \b{\Theta}_i = \b{\Theta}_i^\top \b{\Delta}^\top \b{\Delta} \b{\Theta}_i = (\b{\Delta} \b{\Theta}_i)^\top (\b{\Delta} \b{\Theta}_i) = \b{I}$.
In Eq. (\ref{equation_optimization_kernel_ISCA_allBlocks}), if we embed the constraint in the objective function \cite{boyd2011distributed}, we have:
\begin{equation}\label{equation_ADMM_problem_kernel}
\begin{aligned}
& \underset{\b{\Theta}_i, \b{V}_i \in \mathbb{R}^{n \times p}}{\text{minimize}}
& & \sum_{i=1}^b \big( f(\b{\Theta}_i) + h(\b{\Delta} \b{V}_i) \big), \\
& \text{subject to}
& & \b{\Theta} - \b{V} = \b{0}, 
\end{aligned}
\end{equation}
where $h(\b{\Delta} \b{V}_i) = \mathbb{I}\big((\b{\Delta} \b{V}_i)^\top (\b{\Delta} \b{V}_i) = \b{I}\big)$ and $\b{\Theta} := \cup_{i=1}^b \b{\Theta}_i$ and $\b{V} := \cup_{i=1}^b \b{V}_i$.
Taking $\mathbb{R}^{n \times p} \ni \b{W}_i := \b{\Delta} \b{V}_i$, we can restate Eq. (\ref{equation_ADMM_problem_kernel}) as: $\underset{\b{\Theta}_i, \b{W}_i}{\text{minimize}} \sum_{i=1}^b \big( f(\b{\Theta}_i) + h(\b{W}_i) \big)$, subject to $\b{\Delta}\b{\Theta} - \b{W} = \b{0}$, where $\b{W} := \cup_{i=1}^b \b{W}_i$.
The ADMM solution to this optimization problem is \cite{boyd2011distributed,otero2018alternate}:
\begin{align}
\b{\Theta}_i^{(k+1)} & := \arg \min_{\b{\Theta}_i} \Big( f(\b{\Theta}_i) + (\rho/2)\, ||\b{\Delta}\b{\Theta}_i - \b{W}_i^{(k)} + \b{J}_i^{(k)}||_F^2 \Big), \label{equation_ADMM_Theta_update_kernel} \\
\b{W}_i^{(k+1)} & := \arg \min_{\b{W}_i} \Big( h(\b{W}_i) + (\rho/2)\, ||\b{\Delta}\b{\Theta}_i^{(k+1)} - \b{W}_i + \b{J}_i^{(k)}||_F^2 \Big), \label{equation_ADMM_W_update_kernel} \\
\b{J}^{(k+1)} & := \b{J}^{(k)} + \b{\Delta}\b{\Theta}^{(k+1)} - \b{W}^{(k+1)}. \label{equation_ADMM_J_update_kernel}
\end{align}
With the similar explanations which we had for Eq. (\ref{equation_ADMM_update_2}), we have:
\begin{equation}\label{equation_ADMM_update_kernel_2}
\begin{aligned}
\b{\Theta}_i^{(k+1)} & := \b{\Theta}_i^{(k)} - \eta\, \b{G}(\b{\Theta}_i^{(k)}) - \eta\,\rho\, \b{\Delta}^\top (\b{\Delta}\b{\Theta}_i^{(k)} - \b{W}_i^{(k)} + \b{J}_i^{(k)}), \\
\b{W}_i^{(k+1)} & := \b{Q}_i\,\, \textbf{diag}\Big(\textbf{prox}_{\rho, h}\big(\sigma(\b{\Delta}\b{\Theta}_i^{(k+1)} + \b{J}_i^{(k)})\big)\Big)\,\, \b{\Omega}_i^\top,  \\
\b{J}^{(k+1)} & := \b{J}^{(k)} + \b{\Delta}\b{\Theta}^{(k+1)} - \b{W}^{(k+1)},
\end{aligned}
\end{equation}
where columns of $\b{Q}_i \in \mathbb{R}^{q \times p}$ and $\b{\Omega}_i \in \mathbb{R}^{p \times p}$ are the left and right singular vectors of $(\b{\Delta}\b{\Theta}_i^{(k+1)} + \b{J}_i^{(k)})$.
Iteratively solving Eq. (\ref{equation_ADMM_update_kernel_2}) until convergence gives us the $\b{\Theta}_i$ for for the image blocks indexed by $i$. The $p$ columns of $\b{\Theta}_i$ are the bases for the \textit{kernel ISCA subspace} of the $i$-th block. 
The $i$-th projected block is $\b{\Theta}_i^\top \b{k}_i \in \mathbb{R}^p$ and its dimensions are the \textit{kernel image structural components}. 
Note that $\b{k}_i$, whether it is the kernel over a block in a training image or an out-of-sample image, is normalized and centered. 
Also, note that we had considered the blocks of only one image for Eq. (\ref{equation_ADMM_update_kernel_2}).
Again, with the auto-encoder approach, we can solve these equations in successive epochs in order to find the $b$ subspaces for all the $n$ training images.

\begin{figure}[!t]
\centering
\includegraphics[width=4.7in]{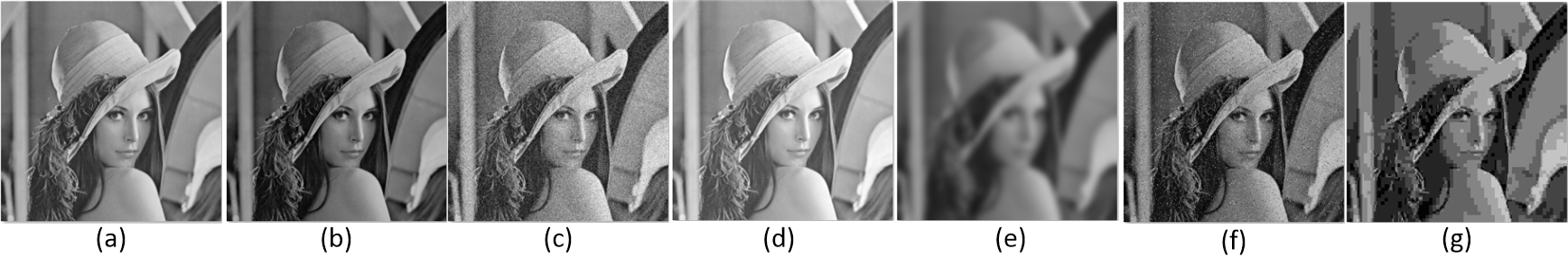}
\caption{Examples from the training dataset: (a) original image, (b) contrast stretched, (c) Gaussian noise, (d) luminance enhanced, (e) Gaussian blurring, (f) salt \& pepper impulse noise, and (g) JPEG distortion.}
\label{fig_training_dataset}
\end{figure}

\begin{figure}[!t]
\centering
\includegraphics[width=3.5in]{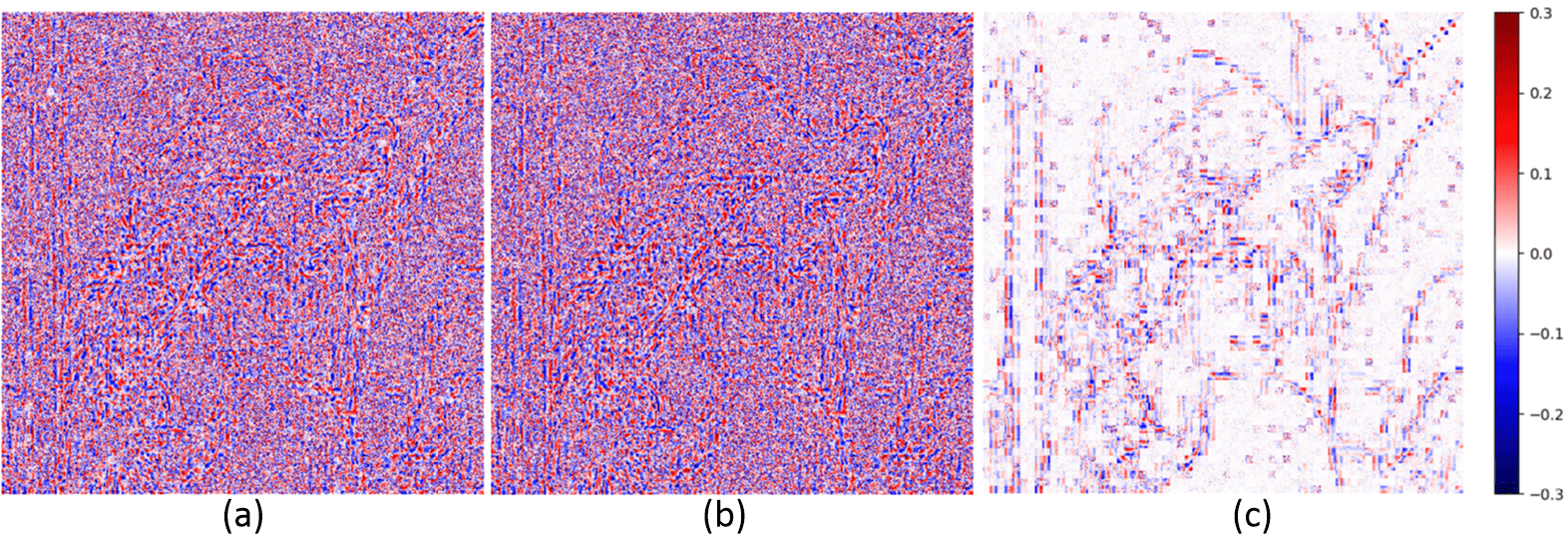}
\caption{The first dimension of the trained (a) $\b{U}$, (b) $\b{V}$, and (c) $\b{J}$ for ISCA.}
\label{fig_projection_directions}
\end{figure}

\section{Experiments}\label{sec:experiments}

\textbf{Training Dataset:}
We formed a dataset out of the standard \textit{Lena} image. Six different types of distortions were applied on the original \textit{Lena} image (see Fig. \ref{fig_training_dataset}), each of which has $20$ images in the dataset with different MSE values. Therefore, the size of the training set is $121$ including the original image. 
For every type of distortion, $20$ different levels of MSE, i.e., from $\text{MSE} = 45$ to $\text{MSE} = 900$ with step $45$, were generated to have images on the equal-MSE or \textit{iso-error} hypersphere \cite{wang2006modern}.

\BlankLine
\noindent
\textbf{Training:}
In our experiments for ISCA, the parameters used were $\rho = 1$ and $\eta=0.1$, and for kernel ISCA, we used $\rho = 0.1$ and $\eta=0.1$. We took $q= 64$ ($8 \times 8$ blocks inspired by \cite{otero2014unconstrained,otero2018alternate}), $p=4$, and $d=512\times 512 = 262144$.
One of the dimensions of the trained $\b{U} = \cup_{i=1}^b \b{U}_i$, $\b{V} = \cup_{i=1}^b \b{V}_i$, and $\b{J} = \cup_{i=1}^b \b{J}_i$ for ISCA are shown in Fig. \ref{fig_projection_directions}. The dual variable $\b{J}$ has captured the edges because edges carry much of the structure information. As expected, $\b{U}$ and $\b{V}$ are close (\textit{Lena} can be seen in them by noticing scrupulously). Note that the variables in kernel ISCA are not $q$-dimensional and thus cannot be displayed in image form. 

\begin{figure}[!t]
\centering
\includegraphics[width=4.8in]{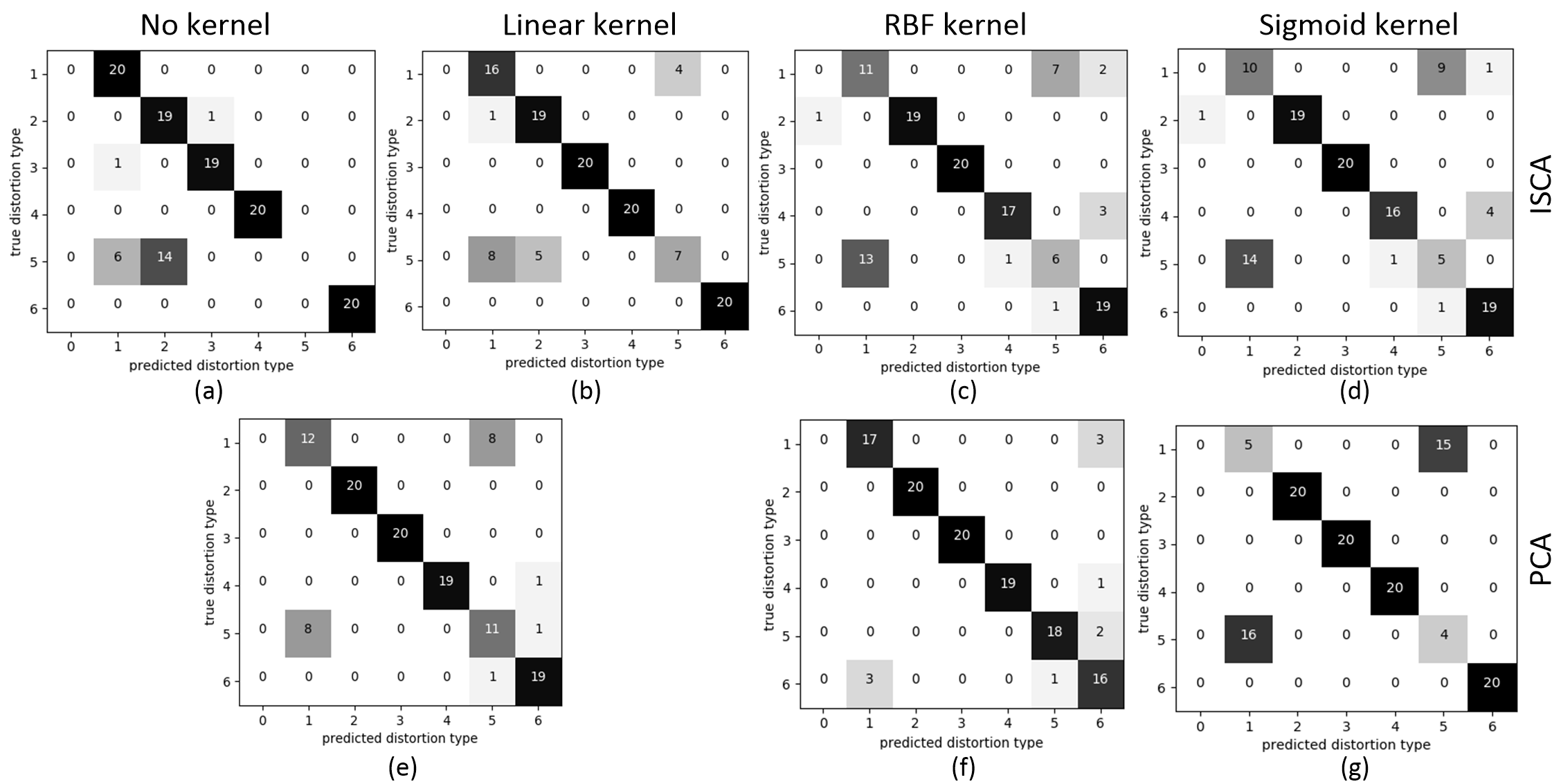}
\caption{Confusion matrices for recognition of distortion types with a 1NN classifier used in the subspace. Matrices (a) and (e) correspond to ISCA and PCA (or linear-kernel PCA), respectively. Matrices (b) to (d) are for kernel ISCA with linear, RBF, and sigmoid kernels. Matrices (f) and (g) are for kernel PCA with RBF and sigmoid kernels. The $0$ label in matrices correspond to the original image and the labels $1$ to $6$ are the distortion types with the same order as in Fig. \ref{fig_training_dataset}.}
\label{fig_confusion_matrices}
\end{figure}

\begin{figure}[!t]
\centering
\includegraphics[width=3.5in]{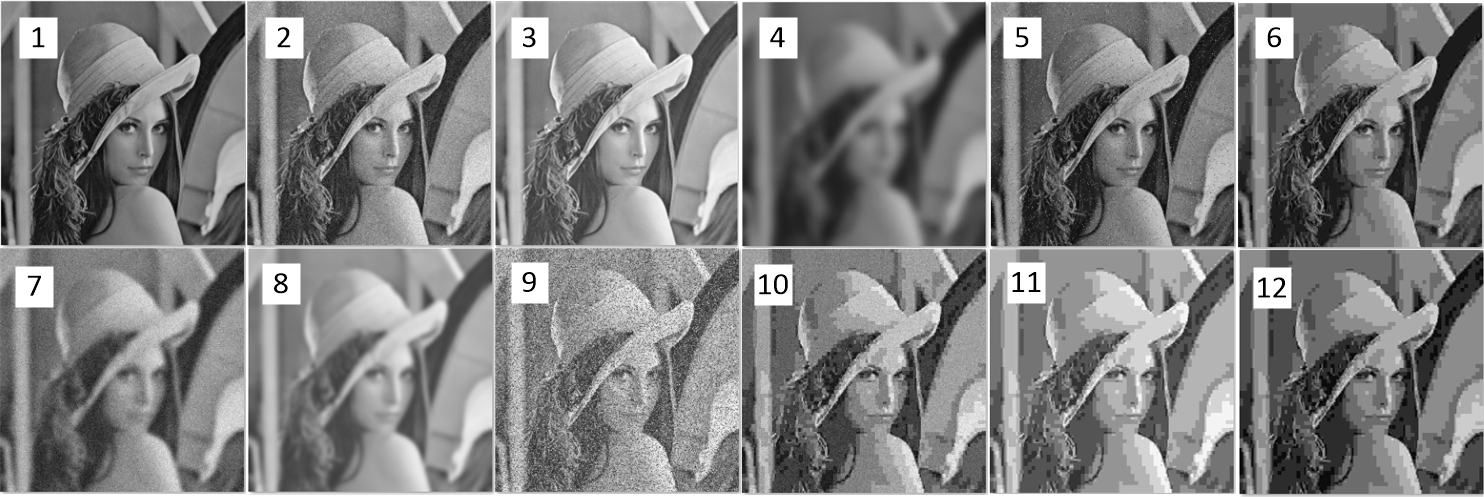}
\caption{Out-of-sample images with different types of distortions having $\text{MSE}=500$: (1) stretching contrast, (2) Gaussian noise, (3) luminance enhancement, (4) Gaussian blurring, (5) impulse noise, (6) JPEG distortion, (7) Gaussian blurring $+$ Gaussian noise, (8) Gaussian blurring $+$ luminance enhancement, (9) impulse noise $+$ luminance enhancement, (10) JPEG distortion $+$ Gaussian noise, (11) JPEG distortion $+$ luminance enhancement, and (12) JPEG distortion $+$ stretching contrast.}
\label{fig_test_dataset}
\end{figure}

\BlankLine
\noindent
\textbf{Projections and Comparisons:}
In order to evaulate the trained ISCA and kernel ISCA subspaces, we projected the training images onto these subspaces. For projecting an image, each of its blocks is projected onto the subspace of that block. 
After projecting all the images, we used the 1-Nearest Neighbor (1NN) classifier to recognize the distortion type of every block. The 1NN is useful to evaluate the subspace by closeness of the projected distortions.  
The distortion type of an image comes from a majority vote among the blocks. 
The linear, Radial Basis function (RBF), and sigmoid kernels were tested for kernel ISCA.
The confusion matrices for distortion recognition are shown in Fig. \ref{fig_confusion_matrices}. Mostly kernel ISCA performed better than ISCA because it works in feature space; although, ISCA performed better for some distortions like contrast stretching and blurring. 
Moreover, we compared with PCA and kernel PCA. PCA showed weakness in contrast stretching. RBF and sigmoid kernels in kernel PCA do not perform well for JPEG distortion and contrast stretching, respectively.

\BlankLine
\noindent
\textbf{Out-of-sample Projections:}
For out-of-sample projection, we created $12$ test images with $\text{MSE}=500$ having different distortions and some having a combination of different distortions (see Fig. \ref{fig_test_dataset}).
We did the same 1NN classification for these images. Table \ref{table_outOfSample_recongition} reports the top two votes of blocks for every image with the percentage of blocks voting for those distortions. ISCA did not recognize luminance enhancement well enough because, for Eq. (\ref{equation_SSIM_distance}), the block is centered while in kernel ISCA, the block is centered in feature space. Overall, both ISCA and kernel ISCA performed very compelling even in recognizing the combination of distortions.

\begin{table*}[!t]
\begin{minipage}{\textwidth}
\caption{Recognition of distortions for out-of-sample images. Letters O, C, G, L, B, I, and J correspond to original image, contrast stretch, Gaussian noise, luminance enhanced, blurring, impulse noise, and JPEG distortion, respectively.}
\label{table_outOfSample_recongition}
\setlength\extrarowheight{5pt}
\centering
\scalebox{0.6}{    
\begin{tabular}{l || c | c | c | c | c | c | c | c | c | c | c | c}
image &  1 & 2 & 3 & 4 & 5 &  6 & 7 & 8 & 9 & 10 & 11 & 12 \\
\hline
\hline
distortion & C & G & L & B & I & J & B $+$ G & B $+$ L & I $+$ L & J $+$ G & J $+$ L & J $+$ C \\
\hline
\hline
\multirow{2}{*}{ISCA}  
& 69.3\% O & 49.1\% G & 69.7\% O & 99.8\% B & 30.3\% G & 96.4\% J & 55.2\% B & 98.7\% B & 48.9\% G & 39.4\% J & 96.4\% J & 97.9\% J \\
& 30.2\% C & 27.2\% I & 29.6\% C & 0.2\% J & 23.8\% I & 3.6\% B & 19.9\% G & 1.3\% J & 33.3\% I & 32.9\% B & 3.6\% B & 2.1\% B \\
\hline
\multirow{2}{*}{kernel ISCA (linear)} 
& 88.1\% C & 59.2\% G & 99.8\% L & 95.9\% B & 37.4\% G & 80.4\% J & 40.8\% G & 93.4\% B & 45.6\% I & 38.7\% G & 70.2\% J & 74.1\% J \\
& 11.2\% I & 25.2\% I & 0.1\% O & 3.4\% J & 32.3\% I & 17.6\% B & 33.4\% B & 5.8\% J & 39.4\% G & 21.7\% J & 27.0\% B & 25.0\% B \\
\hline
\multirow{2}{*}{kernel ISCA (RBF)} 
& 72.0\% C & 79.1\% G & 99.2\% L & 70.6\% B & 39.6\% I & 74.3\% J & 44.8\% G & 88.1\% L & 48.2\% L & 33.1\% G & 82.8\% L & 43.1\% J \\
& 10.9\% I & 5.0\% B & 0.5\% G & 13.1\% C & 36.4\% C & 13.5\% C & 28.5\% B & 6.6\% B & 37.7\% G & 21.6\% L & 8.9\% G & 30.0\% B \\
\hline
\multirow{2}{*}{kernel ISCA (sigmoid)} 
& 80.3\% C & 76.3\% G & 99.6\% L & 76.2\% B & 38.5\% I & 79.3\% J & 47.9\% G & 81.7\% L & 52.1\% L & 37.9\% G & 80.7\% L & 43.9\% J \\
& 7.6\% I & 6.8\% I & 0.2\% G,B & 10.6\% J & 36.5\% C & 10.6\% C & 24.6\% B & 9.8\% B & 26.6\% G & 19.8\% L & 11.3\% G & 31.0\% B \\
\hline
\hline
\end{tabular}%
}
\end{minipage}
\end{table*}

\begin{figure}[!t]
\centering
\includegraphics[width=3.5in]{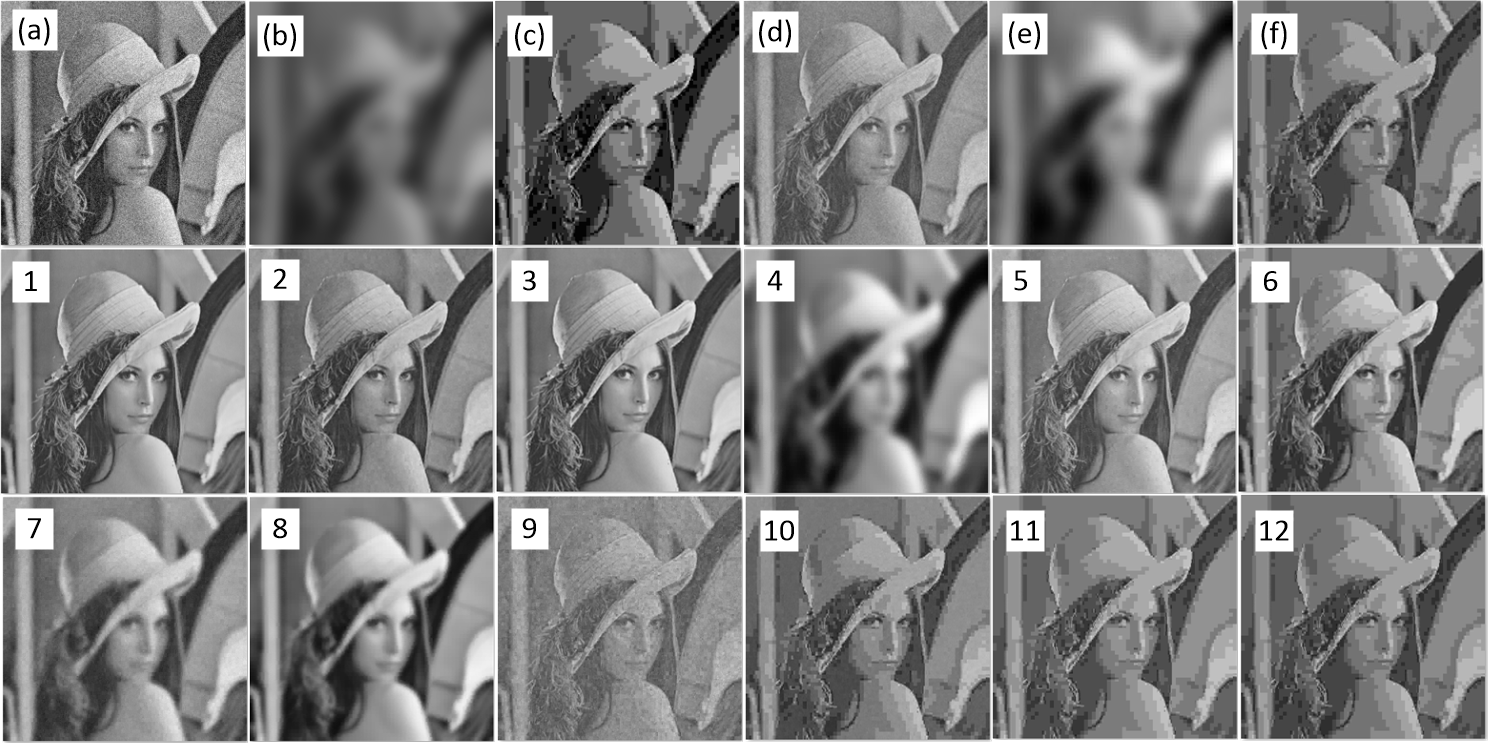}
\caption{Reconstruction of images in ISCA. Reconstruction of the training images (a), (b), and (c) are shown in (d), (e), and (f), respectively. The reconstruction of out-of-sample images shown in Fig. \ref{fig_test_dataset} are shown in the second and third rows.}
\label{fig_reconstructed}
\end{figure}

\BlankLine
\noindent
\textbf{Reconstruction:}
The images can be reconstructed after the projection onto the ISCA subspace. For reconstruction, every block is reconstructed as $\b{U}_i\b{U}_i^\top\breve{\b{x}}_i \in \mathbb{R}^q$ where the mean of block should be added to the reconstruction. Similar to kernel PCA, reconstruction cannot be done in kernel ISCA because $\b{\Theta}_i\, \b{\Theta}_i^\top \b{k}_i \in \mathbb{R}^n \neq \mathbb{R}^q$. Figure \ref{fig_reconstructed} shows reconstruction of some of training and out-of-sample images. As expected, the reconstructed images, for both training and out-of-sample images, are very similar to the original images. 

\section{Conclusion and Future Direction}
This paper introduces the concept of an image structure subspace which captures the structure of an image and discriminates the distortion types.
We hope this will open a broad new field for research in this area and build a greatly needed bridge between the worlds of image quality assessment and manifold learning.

For image structure subspace learning, ISCA and kernel ISCA were proposed, taking inspiration from PCA.
As future work, we can consider designing deeper auto-encoder \cite{goodfellow2016deep} with non-linear activation functions for image structure subspace learning.

\bibliographystyle{splncs}      
\bibliography{references.bib}            

\section{Supplementary Material}

This section is the supplementary material for the paper ``Principal Component Analysis Using Structural Similarity Index for Images''. 
In this paper, the derivation of the mathematical expressions, which were not completely detailed in the main paper, are explained.
We explain the derivation of Eqs. (\ref{equation_f_U}), (\ref{equation_gradient_of_f}), (\ref{equation_ADMM_update_2}), (\ref{equation_f_kernel}), (\ref{equation_gradient_of_f_kernel}), and (\ref{equation_ADMM_update_kernel_2}).

\subsection{Derivation of Eq. (\ref{equation_f_U})}

In the following, we mention the derivation of Eq. (\ref{equation_f_U}):
\begin{align*}
f(\b{U}) := ||\breve{\b{x}} - \b{U}\b{U}^\top\breve{\b{x}}||_S \overset{(a)}{=} \frac{||\breve{\b{x}} - \b{U}\b{U}^\top\breve{\b{x}}||_2^2}{||\breve{\b{x}}||_2^2 + ||\b{U}\b{U}^\top\breve{\b{x}}||_2^2 + c},
\end{align*}
where $(a)$ is because of Eq. (\ref{equation_SSIM_distance}).
The numerator of $f(\b{U})$ is simplified as:
\begin{align*}
||\breve{\b{x}} - \b{U}\b{U}^\top\breve{\b{x}}||_2^2 &= (\breve{\b{x}} - \b{U}\b{U}^\top\breve{\b{x}})^\top (\breve{\b{x}} - \b{U}\b{U}^\top\breve{\b{x}}) = (\breve{\b{x}}^\top - \breve{\b{x}}^\top\b{U}\b{U}^\top) (\breve{\b{x}} - \b{U}\b{U}^\top\breve{\b{x}}) \\
&= \breve{\b{x}}^\top \breve{\b{x}} - \breve{\b{x}}^\top \b{U}\b{U}^\top\breve{\b{x}} - \breve{\b{x}}^\top \b{U}\b{U}^\top\breve{\b{x}} + \breve{\b{x}}^\top \b{U}\underbrace{\b{U}^\top\b{U}}_{\b{I}}\b{U}^\top\breve{\b{x}} \\
& \overset{(a)}{=} \breve{\b{x}}^\top \breve{\b{x}} -2\, \breve{\b{x}}^\top \b{U}\b{U}^\top\breve{\b{x}} + \breve{\b{x}}^\top \b{U}\b{U}^\top\breve{\b{x}} \\
& = \breve{\b{x}}^\top \breve{\b{x}} - \breve{\b{x}}^\top \b{U}\b{U}^\top\breve{\b{x}} = \breve{\b{x}}^\top (\b{I} - \b{U}\b{U}^\top)\,\breve{\b{x}},
\end{align*}
where $(a)$ is because of the constraint $\b{U}^\top \b{U} = \b{I}$ in Eq. (\ref{equation_optimization_ISCA_oneBlock}).

The first term in denominator of $f(\b{U})$ is simplified as:
\begin{align*}
||\breve{\b{x}}||_2^2 = \breve{\b{x}}^\top \breve{\b{x}},
\end{align*}
and the second term in denominator of $f(\b{U})$ is simplified as:
\begin{align*}
||\b{U}\b{U}^\top\breve{\b{x}}||_2^2 &= (\b{U}\b{U}^\top\breve{\b{x}})^\top (\b{U}\b{U}^\top\breve{\b{x}}) = (\breve{\b{x}}^\top\b{U}\b{U}^\top) (\b{U}\b{U}^\top\breve{\b{x}}) \\
& = \breve{\b{x}}^\top\b{U}\underbrace{\b{U}^\top \b{U}}_{\b{I}}\b{U}^\top\breve{\b{x}} \overset{(a)}{=} \breve{\b{x}}^\top\b{U}\b{U}^\top\breve{\b{x}},
\end{align*}
where $(a)$ is because of the constraint $\b{U}^\top \b{U} = \b{I}$ in Eq. (\ref{equation_optimization_ISCA_oneBlock}).
Thus, the denominator of $f(\b{U})$ is:
\begin{align*}
||\breve{\b{x}}||_2^2 + ||\b{U}\b{U}^\top\breve{\b{x}}||_2^2 + c = \breve{\b{x}}^\top \breve{\b{x}} + \breve{\b{x}}^\top\b{U}\b{U}^\top\breve{\b{x}} + c = \breve{\b{x}}^\top (\b{I} + \b{U}\b{U}^\top)\, \breve{\b{x}} + c.
\end{align*}
Therefore, the Eq. (\ref{equation_f_U}) becomes:
\begin{align}
\mathbb{R} \ni f(\b{U}) = \frac{\breve{\b{x}}^\top (\b{I} - \b{U}\b{U}^\top)\, \breve{\b{x}}}{\breve{\b{x}}^\top (\b{I} + \b{U}\b{U}^\top)\, \breve{\b{x}} + c}.
\end{align}

\subsection{Derivation of Eq. (\ref{equation_gradient_of_f})}

In the following, we mention the derivation of Eq. (\ref{equation_gradient_of_f}).
We take the numerator and denominator of derivative of $f(\b{U})$ as:
\begin{alignat*}{2}
& \mathbb{R} \ni \alpha := \breve{\b{x}}^\top (\b{I} - \b{U}\b{U}^\top)\, \breve{\b{x}} &&= \breve{\b{x}}^\top \breve{\b{x}} - \breve{\b{x}}^\top \b{U}\b{U}^\top \breve{\b{x}} \overset{(a)}{=} \breve{\b{x}}^\top \breve{\b{x}} - \textbf{tr}(\breve{\b{x}}^\top \b{U}\b{U}^\top \breve{\b{x}}) \\
& &&\overset{(b)}{=} \breve{\b{x}}^\top \breve{\b{x}} - \textbf{tr}(\b{U}^\top \breve{\b{x}}\breve{\b{x}}^\top \b{U}), \\
& \mathbb{R} \ni \beta := \breve{\b{x}}^\top (\b{I} + \b{U}\b{U}^\top)\, \breve{\b{x}} + c &&= \breve{\b{x}}^\top \breve{\b{x}} + \breve{\b{x}}^\top \b{U}\b{U}^\top \breve{\b{x}} + c \\
& &&\overset{(a)}{=} \breve{\b{x}}^\top \breve{\b{x}} + \textbf{tr}(\breve{\b{x}}^\top \b{U}\b{U}^\top \breve{\b{x}}) + c \\
& &&\overset{(b)}{=} \breve{\b{x}}^\top \breve{\b{x}} + \textbf{tr}(\b{U}^\top \breve{\b{x}}\breve{\b{x}}^\top \b{U}) + c,
\end{alignat*}
to have $f(\b{U}) = \alpha / \beta$, where $\textbf{tr}(.)$ denotes the trace of matrix, $(a)$ is because a scalar is equal to its trace, and $(b)$ is because of the cyclic property of trace.

The derivative of $f(\b{U})$ with respect to $\b{U}$ is:
\begin{align*}
\mathbb{R}^{q \times p} \ni \frac{\partial f(\b{U})}{\partial\, \b{U}} &= \frac{1}{\beta^2}\Big[(\beta)(-2\, \breve{\b{x}}\breve{\b{x}}^\top \b{U}) - (\alpha) (2\, \breve{\b{x}}\breve{\b{x}}^\top \b{U}) \Big] \\
&= \frac{2}{\beta} (-1 - \frac{\alpha}{\beta})\, \breve{\b{x}}\breve{\b{x}}^\top \b{U} = \frac{-2}{\beta} (1 + f(\b{U}))\, \breve{\b{x}}\breve{\b{x}}^\top \b{U} \\
&= \frac{-2\, (1 + f(\b{U}))}{\beta}\, \breve{\b{x}}\breve{\b{x}}^\top \b{U}.
\end{align*}
Note that $\beta = \breve{\b{x}}^\top (\b{I} + \b{U}\b{U}^\top)\, \breve{\b{x}} + c = ||\breve{\b{x}}||_2^2 + ||\b{U}\b{U}^\top\breve{\b{x}}||_2^2 + c$.
Therefore, the gradient of $f(\b{U})$ is obtained:
\begin{align}
\mathbb{R}^{q \times p} \ni \b{G}(\b{U}) := \frac{\partial f(\b{U})}{\partial\, \b{U}} = \frac{-2\,(1 + f(\b{U}))}{||\breve{\b{x}}||_2^2 + ||\b{U}\b{U}^\top\breve{\b{x}}||_2^2 + c} \,\breve{\b{x}}\breve{\b{x}}^\top\b{U}.
\end{align}

\subsection{Derivation of Update of $\b{U}_i$ in Eq. (\ref{equation_ADMM_update_2})}

In the following, we mention the derivation of Eq. (\ref{equation_ADMM_update_2}).
The Eq. (\ref{equation_ADMM_U_update}) is:
\begin{align*}
\b{U}_i^{(k+1)} := \arg \min_{\b{U}_i} \Big( f(\b{U}_i) + (\rho/2)\, ||\b{U}_i - \b{V}_i^{(k)} + \b{J}_i^{(k)}||_F^2 \Big).
\end{align*}
The objective function can be simplified as:
\begin{align*}
f(\b{U}_i) &+ (\rho/2)\, ||\b{U}_i - \b{V}_i^{(k)} + \b{J}_i^{(k)}||_F^2 \\
&= f(\b{U}_i) + (\rho/2)\, \textbf{tr} \Big( (\b{U}_i - \b{V}_i^{(k)} + \b{J}_i^{(k)})^\top (\b{U}_i - \b{V}_i^{(k)} + \b{J}_i^{(k)}) \Big) \\
&= f(\b{U}_i) + (\rho/2)\, \textbf{tr} \Big( (\b{U}_i^\top - \b{V}_i^{(k)\top} + \b{J}_i^{(k)\top}) (\b{U}_i - \b{V}_i^{(k)} + \b{J}_i^{(k)}) \Big) \\
&= f(\b{U}_i) + (\rho/2)\, \textbf{tr} \Big( \b{U}_i^\top \b{U}_i - \b{U}_i^\top \b{V}_i^{(k)} + \b{U}_i^\top \b{J}_i^{(k)} - \b{V}_i^{(k)\top} \b{U}_i \\
&+ \b{V}_i^{(k)\top} \b{V}_i^{(k)} - \b{V}_i^{(k)\top} \b{J}_i^{(k)} + \b{J}_i^{(k)} \b{U}_i - \b{J}_i^{(k)} \b{V}_i^{(k)} + \b{J}_i^{(k)\top} \b{J}_i^{(k)} \Big).
\end{align*}
The gradient of the objective function with respect to $\b{U}_i$ is:
\begin{align*}
&\frac{\partial}{\partial \b{U}_i}\Big( f(\b{U}_i) + (\rho/2)\, ||\b{U}_i - \b{V}_i^{(k)} + \b{J}_i^{(k)}||_F^2 \Big) \\
&= \b{G}(\b{U}_i) + (\rho/2)\, \Big( 2\,\b{U}_i - \b{V}_i^{(k)} + \b{J}_i^{(k)} - \b{V}_i^{(k)} + \b{J}_i^{(k)} \Big) \\
&= \b{G}(\b{U}_i) + \rho\, ( \b{U}_i - \b{V}_i^{(k)} + \b{J}_i^{(k)} ).
\end{align*}
Therefore, the iteration in gradient descent is:
\begin{align}
\b{U}_i^{(k+1)} & := \b{U}_i^{(k)} - \eta\, \frac{\partial}{\partial \b{U}_i}( ... ) = \b{U}_i^{(k)} - \eta\, \b{G}(\b{U}_i^{(k)}) - \eta\,\rho\, (\b{U}_i^{(k)} - \b{V}_i^{(k)} + \b{J}_i^{(k)}),
\end{align}
where $\eta$ is the learning rate and $\frac{\partial}{\partial \b{U}_i}( ... )$ is derivative of the objective function.

\subsection{Derivation of Eq. (\ref{equation_f_kernel})}

In the following, we mention the derivation of Eq. (\ref{equation_f_kernel}):
\begin{align*}
f(\b{\Theta}_i) &= ||\b{\phi}(\breve{\b{x}}_i) - \b{\Phi}(\breve{\b{X}}_i)\, \b{\Theta}_i\, \b{\Theta}_i^\top \b{k}_i||_S \\
&= ||\b{\phi}(\breve{\b{x}}_i) - \b{\Phi}(\breve{\b{X}}_i)\, \b{\Theta}_i\, \b{\Theta}_i^\top \b{\Phi}(\breve{\b{X}}_i)^\top \b{\phi}(\breve{\b{x}}_i)||_S \\
&\overset{(a)}{=} \frac{||\b{\phi}(\breve{\b{x}}_i) - \b{\Phi}(\breve{\b{X}}_i)\, \b{\Theta}_i\, \b{\Theta}_i^\top \b{\Phi}(\breve{\b{X}}_i)^\top \b{\phi}(\breve{\b{x}}_i)||_2^2}{||\b{\phi}(\breve{\b{x}}_i)||_2^2 + ||\b{\Phi}(\breve{\b{X}}_i)\, \b{\Theta}_i\, \b{\Theta}_i^\top \b{\Phi}(\breve{\b{X}}_i)^\top \b{\phi}(\breve{\b{x}}_i)||_2^2 + c},
\end{align*}
where $(a)$ is because of Eq. (\ref{equation_SSIM_distance}).
The numerator of $f(\b{\Theta}_i)$ is simplified as:
\begin{align*}
&||\b{\phi}(\breve{\b{x}}_i) - \b{\Phi}(\breve{\b{X}}_i)\, \b{\Theta}_i\, \b{\Theta}_i^\top \b{\Phi}(\breve{\b{X}}_i)^\top \b{\phi}(\breve{\b{x}}_i)||_2^2 \\
&= \big(\b{\phi}(\breve{\b{x}}_i) - \b{\Phi}(\breve{\b{X}}_i)\, \b{\Theta}_i\, \b{\Theta}_i^\top \b{\Phi}(\breve{\b{X}}_i)^\top \b{\phi}(\breve{\b{x}}_i)\big)^\top \\
&~~~~~~~~~~~~~~~~~~~~~~~~~~~~~~~~~~~~~~~~~~~~~~~ \big(\b{\phi}(\breve{\b{x}}_i) - \b{\Phi}(\breve{\b{X}}_i)\, \b{\Theta}_i\, \b{\Theta}_i^\top \b{\Phi}(\breve{\b{X}}_i)^\top \b{\phi}(\breve{\b{x}}_i)\big) \\
&= \big(\b{\phi}(\breve{\b{x}}_i)^\top - \b{\phi}(\breve{\b{x}}_i)^\top \b{\Phi}(\breve{\b{X}}_i)\, \b{\Theta}_i\, \b{\Theta}_i^\top \b{\Phi}(\breve{\b{X}}_i)^\top\big) \\
&~~~~~~~~~~~~~~~~~~~~~~~~~~~~~~~~~~~~~~~~~~~~~~~ \big(\b{\phi}(\breve{\b{x}}_i) - \b{\Phi}(\breve{\b{X}}_i)\, \b{\Theta}_i\, \b{\Theta}_i^\top \b{\Phi}(\breve{\b{X}}_i)^\top \b{\phi}(\breve{\b{x}}_i)\big) \\
&= \underbrace{\b{\phi}(\breve{\b{x}}_i)^\top \b{\phi}(\breve{\b{x}}_i)}_{k_i} - \underbrace{\b{\phi}(\breve{\b{x}}_i)^\top \b{\Phi}(\breve{\b{X}}_i)}_{\b{k}_i^\top}\, \b{\Theta}_i\, \b{\Theta}_i^\top \underbrace{\b{\Phi}(\breve{\b{X}}_i)^\top \b{\phi}(\breve{\b{x}}_i)}_{\b{k}_i} \\
&~~~~~~~~~~~~~~~~~~~ -\underbrace{\b{\phi}(\breve{\b{x}}_i)^\top \b{\Phi}(\breve{\b{X}}_i)}_{\b{k}_i^\top}\, \b{\Theta}_i\, \b{\Theta}_i^\top \underbrace{\b{\Phi}(\breve{\b{X}}_i)^\top \b{\phi}(\breve{\b{x}}_i)}_{\b{k}_i} \\
&~~~~~~~~~~~~~~~~~~~ + \underbrace{\b{\phi}(\breve{\b{x}}_i)^\top \b{\Phi}(\breve{\b{X}}_i)}_{\b{k}_i^\top}\, \b{\Theta}_i\, \b{\Theta}_i^\top \underbrace{\b{\Phi}(\breve{\b{X}}_i)^\top \b{\Phi}(\breve{\b{X}}_i)}_{\b{K}_i}\, \b{\Theta}_i\, \b{\Theta}_i^\top \underbrace{\b{\Phi}(\breve{\b{X}}_i)^\top \b{\phi}(\breve{\b{x}}_i)}_{\b{k}_i} \\
&= k_i - \b{k}_i^\top \b{\Theta}_i\, \b{\Theta}_i^\top \b{k}_i - \b{k}_i^\top \b{\Theta}_i\, \b{\Theta}_i^\top \b{k}_i + \b{k}_i^\top \b{\Theta}_i\, \underbrace{\b{\Theta}_i^\top \b{K}_i\, \b{\Theta}_i}_{\b{I}}\, \b{\Theta}_i^\top \b{k}_i \\
&\overset{(a)}{=} k_i - \b{k}_i^\top \b{\Theta}_i\, \b{\Theta}_i^\top \b{k}_i - \b{k}_i^\top \b{\Theta}_i\, \b{\Theta}_i^\top \b{k}_i + \b{k}_i^\top \b{\Theta}_i\, \b{\Theta}_i^\top \b{k}_i = k_i - \b{k}_i^\top \b{\Theta}_i\, \b{\Theta}_i^\top \b{k}_i,
\end{align*}
where $(a)$ is because of the constraint $\b{\Theta}_i^\top \b{K}_i\, \b{\Theta}_i = \b{I}$ in Eq. (\ref{equation_optimization_kernel_ISCA_allBlocks}).

The first term in denominator of $f(\b{\Theta}_i)$ is simplified as:
\begin{align*}
||\b{\phi}(\breve{\b{x}}_i)||_2^2 = \b{\phi}(\breve{\b{x}}_i)^\top \b{\phi}(\breve{\b{x}}_i) = k_i,
\end{align*}
and the second term in denominator of $f(\b{\Theta}_i)$ is simplified as:
\begin{align*}
&||\b{\Phi}(\breve{\b{X}}_i)\, \b{\Theta}_i\, \b{\Theta}_i^\top \b{\Phi}(\breve{\b{X}}_i)^\top \b{\phi}(\breve{\b{x}}_i)||_2^2 \\
&= \big(\b{\Phi}(\breve{\b{X}}_i)\, \b{\Theta}_i\, \b{\Theta}_i^\top \b{\Phi}(\breve{\b{X}}_i)^\top \b{\phi}(\breve{\b{x}}_i)\big)^\top \big(\b{\Phi}(\breve{\b{X}}_i)\, \b{\Theta}_i\, \b{\Theta}_i^\top \b{\Phi}(\breve{\b{X}}_i)^\top \b{\phi}(\breve{\b{x}}_i)\big) \\
&= \big( \b{\phi}(\breve{\b{x}}_i)^\top \b{\Phi}(\breve{\b{X}}_i) \b{\Theta}_i\, \b{\Theta}_i^\top \b{\Phi}(\breve{\b{X}}_i)^\top \big) \big(\b{\Phi}(\breve{\b{X}}_i)\, \b{\Theta}_i\, \b{\Theta}_i^\top \b{\Phi}(\breve{\b{X}}_i)^\top \b{\phi}(\breve{\b{x}}_i)\big) \\
&= \underbrace{\b{\phi}(\breve{\b{x}}_i)^\top \b{\Phi}(\breve{\b{X}}_i)}_{\b{k}_i^\top} \b{\Theta}_i\, \b{\Theta}_i^\top \underbrace{\b{\Phi}(\breve{\b{X}}_i)^\top \b{\Phi}(\breve{\b{X}}_i)}_{\b{K}_i}\, \b{\Theta}_i\, \b{\Theta}_i^\top \underbrace{\b{\Phi}(\breve{\b{X}}_i)^\top \b{\phi}(\breve{\b{x}}_i)}_{\b{k}_i} \\
&= \b{k}_i^\top \b{\Theta}_i\, \underbrace{\b{\Theta}_i^\top \b{K}_i\, \b{\Theta}_i}_{\b{I}}\, \b{\Theta}_i^\top \b{k}_i \overset{(a)}{=} \b{k}_i^\top \b{\Theta}_i\, \b{\Theta}_i^\top \b{k}_i,
\end{align*}
where $(a)$ is because of the constraint $\b{\Theta}_i^\top \b{K}_i\, \b{\Theta}_i = \b{I}$ in Eq. (\ref{equation_optimization_kernel_ISCA_allBlocks}).
Therefore, the Eq. (\ref{equation_f_kernel}) is obtained:
\begin{align}
\mathbb{R} \ni f(\b{\Theta}_i) = \frac{k_i - \b{k}_i^\top \b{\Theta}_i \b{\Theta}_i^\top \b{k}_i}{k_i + \b{k}_i^\top \b{\Theta}_i \b{\Theta}_i^\top \b{k}_i + c}.
\end{align}

\subsection{Derivation of Eq. (\ref{equation_gradient_of_f_kernel})}

In the following, we mention the derivation of Eq. (\ref{equation_gradient_of_f_kernel}).
We take the numerator and denominator of derivative of $f(\b{\Theta}_i)$ as:
\begin{align*}
& \mathbb{R} \ni \alpha := k_i - \b{k}_i^\top \b{\Theta}_i \b{\Theta}_i^\top \b{k}_i \overset{(a)}{=} k_i - \textbf{tr}(\b{k}_i^\top \b{\Theta}_i \b{\Theta}_i^\top \b{k}_i) \overset{(b)}{=} k_i - \textbf{tr}(\b{\Theta}_i^\top \b{k}_i\, \b{k}_i^\top \b{\Theta}_i), \\
& \mathbb{R} \ni \beta := k_i + \b{k}_i^\top \b{\Theta}_i \b{\Theta}_i^\top \b{k}_i + c \overset{(a)}{=} k_i + \textbf{tr}(\b{k}_i^\top \b{\Theta}_i \b{\Theta}_i^\top \b{k}_i) + c \\
&~~~~~~~~ \overset{(b)}{=} k_i + \textbf{tr}(\b{\Theta}_i^\top \b{k}_i\, \b{k}_i^\top \b{\Theta}_i) + c,
\end{align*}
to have $f(\b{\Theta}_i) = \alpha / \beta$, where $\textbf{tr}(.)$ denotes the trace of matrix, $(a)$ is because a scalar is equal to its trace, and $(b)$ is because of the cyclic property of trace.

The derivative of $f(\b{\Theta}_i)$ with respect to $\b{\Theta}_i$ is:
\begin{align*}
\mathbb{R}^{n \times p} \ni \frac{\partial f(\b{\Theta}_i)}{\partial\, \b{\Theta}_i} &= \frac{1}{\beta^2}\Big[(\beta)(-2\, \b{k}_i\,\b{k}_i^\top \b{\Theta}_i) - (\alpha) (2\, \b{k}_i\,\b{k}_i^\top \b{\Theta}_i) \Big] \\
&= \frac{2}{\beta} (-1 - \frac{\alpha}{\beta})\, \b{k}_i\,\b{k}_i^\top \b{\Theta}_i = \frac{-2}{\beta} (1 + f(\b{\Theta}_i))\, \b{k}_i\,\b{k}_i^\top \b{\Theta}_i \\
&= \frac{-2\, (1 + f(\b{\Theta}_i))}{\beta}\, \b{k}_i\,\b{k}_i^\top \b{\Theta}_i.
\end{align*}
Therefore, the gradient of $f(\b{\Theta}_i)$ is obtained:
\begin{align}
\mathbb{R}^{n \times p} \ni \b{G}(\b{\Theta}_i) := \frac{\partial f(\b{\Theta}_i)}{\partial\, \b{\Theta}_i} = \frac{-2\,(1 + f(\b{\Theta}_i))}{k_i + \b{k}_i^\top \b{\Theta}_i \b{\Theta}_i^\top \b{k}_i + c} \,\b{k}_i \b{k}_i^\top\b{\Theta}_i.
\end{align}

\subsection{Derivation of Update of $\b{\Theta}_i$ in Eq. (\ref{equation_ADMM_update_kernel_2})}

In the following, we mention the derivation of Eq. (\ref{equation_ADMM_update_kernel_2}).
The Eq. (\ref{equation_ADMM_Theta_update_kernel}) is:
\begin{align*}
\b{\Theta}_i^{(k+1)} & := \arg \min_{\b{\Theta}_i} \Big( f(\b{\Theta}_i) + (\rho/2)\, ||\b{\Delta}\b{\Theta}_i - \b{W}_i^{(k)} + \b{J}_i^{(k)}||_F^2 \Big).
\end{align*}
The objective function can be simplified as:
\begin{align*}
f(\b{\Theta}_i) &+ (\rho/2)\, ||\b{\Delta}\b{\Theta}_i - \b{W}_i^{(k)} + \b{J}_i^{(k)}||_F^2 \\
&= f(\b{\Theta}_i) + (\rho/2)\, \textbf{tr} \Big( (\b{\Delta}\b{\Theta}_i - \b{W}_i^{(k)} + \b{J}_i^{(k)})^\top (\b{\Delta}\b{\Theta}_i - \b{W}_i^{(k)} + \b{J}_i^{(k)}) \Big) \\
&= f(\b{\Theta}_i) + (\rho/2)\, \textbf{tr} \Big( (\b{\Theta}_i^\top\b{\Delta}^\top - \b{W}_i^{(k)\top} + \b{J}_i^{(k)\top}) (\b{\Delta}\b{\Theta}_i - \b{W}_i^{(k)} + \b{J}_i^{(k)}) \Big) \\
&= f(\b{\Theta}_i) + (\rho/2)\, \textbf{tr} \Big( \b{\Theta}_i^\top\b{\Delta}^\top \b{\Delta}\b{\Theta}_i - \b{\Theta}_i^\top\b{\Delta}^\top \b{W}_i^{(k)} + \b{\Theta}_i^\top\b{\Delta}^\top \b{J}_i^{(k)} \\
&- \b{W}_i^{(k)\top} \b{\Delta}\b{\Theta}_i + \b{W}_i^{(k)\top} \b{W}_i^{(k)} - \b{W}_i^{(k)\top} \b{J}_i^{(k)} + \b{J}_i^{(k)\top} \b{\Delta}\b{\Theta}_i \\
&- \b{J}_i^{(k)\top} \b{W}_i^{(k)} + \b{J}_i^{(k)\top} \b{J}_i^{(k)} \Big).
\end{align*}
The gradient of the objective function with respect to $\b{\Theta}_i$ is:
\begin{align*}
&\frac{\partial}{\partial \b{\Theta}_i}\Big( f(\b{\Theta}_i) + (\rho/2)\, ||\b{\Delta}\b{\Theta}_i - \b{W}_i^{(k)} + \b{J}_i^{(k)}||_F^2 \Big) \\
&= \b{G}(\b{\Theta}_i) + (\rho/2)\, ( 2\,\b{\Delta}^\top \b{\Delta}\b{\Theta}_i  - \b{\Delta}^\top \b{W}_i^{(k)} + \b{\Delta}^\top \b{J}_i^{(k)} - \b{\Delta}^\top \b{W}_i^{(k)} + \b{\Delta}^\top \b{J}_i^{(k)} ) \\
&= \b{G}(\b{\Theta}_i) + \rho\, \b{\Delta}^\top ( \b{\Delta}\b{\Theta}_i  - \b{W}_i^{(k)} + \b{J}_i^{(k)} ).
\end{align*}
Therefore, the iteration in gradient descent is:
\begin{align}
\b{\Theta}_i^{(k+1)} & := \b{\Theta}_i^{(k)} - \eta\, \frac{\partial}{\partial \b{\Theta}_i}( ... ) \nonumber \\
&= \b{\Theta}_i^{(k)} - \eta\, \b{G}(\b{\Theta}_i^{(k)}) - \eta\,\rho\, \b{\Delta}^\top (\b{\Delta}\b{\Theta}_i^{(k)} - \b{W}_i^{(k)} + \b{J}_i^{(k)}),
\end{align}
where $\eta$ is the learning rate and $\frac{\partial}{\partial \b{\Theta}_i}( ... )$ is derivative of the objective function.

\end{document}